\documentclass[pre,twocolumn,twoside,showpacs,floatfix]{revtex4}
\usepackage{dcolumn}%
\usepackage{bm}%
\usepackage{graphicx,textcomp}
\usepackage{amssymb,amsfonts,amsmath}
\usepackage{color}

\begin{document}

\title{Coagulation with product kernel and arbitrary initial conditions:\\ Exact kinetics within the Marcus-Lushnikov framework}

\author{Agata Fronczak, Micha\l$\;$\L epek, Pawe\l$\;$Kukli\'nski, Piotr Fronczak}

\affiliation{Faculty of Physics, Warsaw University of Technology, Koszykowa 75, PL-00-662 Warsaw, Poland}

\date{\today}


\begin{abstract}
The time evolution of a system of coagulating particles under the product kernel and arbitrary initial conditions is studied. Using the improved Marcus-Lushnikov approach, the master equation is solved for the probability $W(Q,t)$ to find the system in a given mass spectrum $Q=\{n_1,n_2,\dots,n_g\dots\}$, with $n_g$ being the number of particles of size $g$. The exact expression for the average number of particles, $\langle n_g(t)\rangle$, at arbitrary time $t$ is derived and its validity is confirmed in numerical simulations of several selected initial mass spectra.
\end{abstract}


\maketitle

\section{Introduction}\label{SectIntro}

Coagulation is widespread in nature. In physics and chemistry, it is traditionally mentioned in reference to different polymerization phenomena, and when the process of particles' formation in dispersed media (aerosols and hydrosols) is studied \cite{2006book_Seinfeld, 2000book_Friedlander, 1998book_Meakin}. In this regard, obligatory references include famous contributions made by Smoluchowski \cite{1916_Smoluchowski}, Flory \cite{1941a_Flory, 1941b_Flory, 1941c_Flory} and Stockmayer \cite{1943_Stockmayer} (see also more contemporary papers: \cite{1968_Marcus, 1978_Lushnikov, 1980_Ziff, 1983_Ziff, 1983_Hendriks, 1986_Dongen}, and review articles:~\cite{1999_RevAldous, 2003_PhysRepLeyvraz, 2006_PhysDWattis}). And although various issues related to coagulation began to be studied many decades ago, its full understanding is still far from complete. At the same time, interest in coagulation processes is not weakening at all. This is most likely due to a broad range of interdisciplinary applications of the process, which include: percolation phenomena in random graphs and complex networks \cite{2005_JPhysALushnikov, 2009_ScienceAchlioptas, 2010_PRLCosta, 2010_PRECho, 2016_PRLCho, 2016_Conv}, mathematical population genetics \cite{2005book_Hein}, pattern formation in different social \cite{2014_PREMatsoukas, 2014_SciRepMatsoukas}, biological \cite{2007_EcolModelSaadi} and man-made systems \cite{2002_PREDubovik}, and many others. 

The simplest example of the coagulation process is the evolution of a closed system of particles (clusters) that join irreversibly during binary collisions (so-called coagulation acts), according to the following scheme:
\begin{equation}\label{act0}
(g)+(l)\stackrel{K(g,l)}{\longrightarrow}(g+l),
\end{equation}
where $(g)$ stands for a cluster of mass $g$ (we assume that $g$ is a natural number) and $K(g,l)$ is the coagulation kernel representing rate of the process. Over time, the number of clusters decreases and distribution of their sizes changes. Kinetics of the process strongly depends on $K(g,l)$. In particular, when the system starts to evolve from all clusters having the same size (so-called monodisperse initial conditions), it is well known (see e.g. \cite{2010book_Krapivsky}) that for $K(g,l)$ growing fast enough, at some finite time $t_c$, a giant particle emerges, which brings together a fraction of the mass of the whole system. To distinguish this particle from other particles, usually much smaller, which form the so-called \textit{sol}, this giant particle is called the \textit{gel}. The phenomenon of gel formation is an example of non-equilibrium phase transition. The best-known example of a gelling kernel is the multiplicative (product) kernel: $K(g,l)\propto gl$. The constant kernel, $K(g,l)=const$, and the additive kernel, $K(g,l)\propto g+l$, are examples of non-gelling kernels, in which the sol-gel transition is not observed when monodisperse initial conditions are assumed. The above listed kernels are important, because kinetics of coagulation processes in systems evolving according to these kernels under the simplest (monodisperse) initial conditions, were ``exactly solved'', thus becoming reference models of coagulation. 

In the last sentence, the term in quotes: ``exactly solved'' is of special meaning which needs an explanation. Namely, there are some theoretical approaches to modeling coagulation. The best known approach relies on the famous Smoluchowski equation \cite{1916_Smoluchowski} which constitutes an infinite system of coupled nonlinear differential equations for the average number of clusters of a given size, and provides mean-field (and thus approximate) time evolution of the cluster size distribution. Therefore, explicit solutions to this equation (e.g. \cite{1980_Ziff, 1983_Ziff, 1983_Hendriks, 1986_Dongen, 2012_PRELushnikov, 2013_PRELushnikov}) are not ``exact solutions" of the coagulation process. Its genuine exact solutions (without any approximations) for some particular cases (including the mentioned kernels -~constant, additive and multiplicative~- under monodisperse initial conditions) were obtained through direct counting of system states (see \cite{2018_PREFronczak}), or as solutions to the master equation governing the time evolution of the probability distribution over these states (see \cite{1974_Bayewitz, 1985_Hendriks, 2004_PRLLushnikov, 2005_PRELushnikov, 2011_JPhysALushnikov}). The approach resulting from the master equation was first proposed by Marcus \cite{1968_Marcus} in the late 1960s. In the late 1970s it was used by several researchers \cite{1974_Bayewitz, 1985_Hendriks}, among others by Lushnikov \cite{1978_Lushnikov}, who not so long ago again dealt with coagulation and obtained several significant results using this formalism \cite{2006_PhysDLushnikov, 2011_NucLushnikov}.

The case of initial conditions other than monodisperse is much more complicated and relatively less researched. For example, until quite recently, it was thought that, in the thermodynamic limit, behavior of different coagulating systems (with non-gelling and gelling kernels, before and after the gel time $t_c$) is not sensitive to initial conditions and falls into specific universality classes. These classes were to be characterized by dynamical scaling solutions (i.e. time dependent cluster size distributions) which are similar to the solutions arising from monodisperse initial conditions \cite{2003_PhysRepLeyvraz, 2006_PhysDLeyvraz}. Recent results in this area, however, suggest that the behavior of coagulating systems for arbitrary initial conditions is much more complicated. 

For example, solutions to the Smoluchowski equation for the product kernel and algebraically decaying initial conditions show that there exist three different universality classes which depend on the finiteness of the second and the third moment of the initial mass distribution \cite{2015_PhysALeyvraz}. Another example concerns kernels that are traditionally considered non-gelling (e.g. the additive kernel), and which, according to the Smoluchowski equation, under some initial conditions, may become gelling \cite{2004_Menon}. The above ``examples'' are very interesting but also controversial. Concerns about their validity are justified, especially that, essentially, everything we know about coagulation with arbitrary initial particle mass spectra arises from solutions to the Smoluchowski equation, which are inherently approximate (mean-field). In the theory of equilibrium phase transitions, mean-field solutions do not always give correct results, especially when it comes to the behavior of systems in the vicinity of critical points. 

Being aware of problems that may arise from the Smoluchowski equation, we hope our result - the exact solution of the coagulation process with the product kernel and arbitrary initial conditions - which is presented in this paper, will be of some value for people dealing with the theory of these non-equilibrium phenomena. To obtain the result we refine a bit the prominent solution of finite coagulating systems with product kernel and monodisperse initial conditions, which was given some time ago by Lushnikov \cite{2004_PRLLushnikov, 2005_PRELushnikov}. To this end, we use some ideas and formulas, which originate in combinatorics. More specifically, we use the so-called Bell polynomials, which, although explicitly appear in Lushnikov's papers, were unnoticed there. We show that the mentioned polynomials do not only allow one to get some new results, but they also significantly simplify the whole approach.

The paper is organized as follows. In Sec.~\ref{SecReview} we review the Marcus-Lushnikov approach. The exact result of Lushnikov on coagulation with product kernel for monodisperse initial conditions is also presented in this section. The reader who is familiar with the mentioned results may skip reading this section. However, we encourage to read it because in the next section we refer several times to the various equations included therein. In addition, at the end of Sec.~\ref{SecReview}, we introduce Bell polynomials and discuss their basic properties. In Sec.~\ref{SecAF}, we reformulate the Lushnikov solution with the use of Bell polynomials. The Lushnikov result is then used to obtain the exact solution of the coagulation process under any initial conditions. In this section, the time dependence of the particle mass spectrum is studied (analytically and numerically) for some concrete initial particle mass spectra. Section~\ref{SecSum} concludes the paper.

\section{Review of known results}\label{SecReview}

\subsection{Marcus-Lushnikov approach}

The idea of the Marcus-Lushnikov approach is simple \cite{2006_PhysDLushnikov, 2011_NucLushnikov}. Every single state of the system is described as:
\begin{eqnarray}\label{defQ}
Q\!=\!\{n_1,n_2,\dots,n_g,\dots\}
\end{eqnarray}
where $n_g\geq 0$ is the number of clusters of mass $g$ (so-called $g$-mers), with $g$ being the number of monomeric units. Because the considered system is finite and closed, its total mass, $M$, does not change in time. For this reason, the sequence $\{n_g\}$ in Eq.~(\ref{defQ}) is not arbitrary, but satisfies the following condition:
\begin{equation}\label{ngEq}
\sum_{g=1}^Mg\,n_g=M.
\end{equation}
A single coagulation act, Eq.~(\ref{act0}), changes the proceeding system state, $Q$ (see Eq.~(\ref{defQ})), into the new one, $Q^+$, which is given by
\begin{eqnarray}\label{defQplus1}
Q^+\!=\!\{n_1,\dots,n_g\!-\!1,\dots,n_l\!-\!1,\dots,n_{g+l}\!+\!1,\dots\}, 
\end{eqnarray}
when $g<l$, or  
\begin{eqnarray}\label{defQplus2}
Q^+=\{n_1,\dots,n_g\!-\!2,\dots,n_{2g}\!+\!1,\dots\},
\end{eqnarray}
for $g=l$. Correspondingly, if as a result of the coagulation described by Eq.~(\ref{act0}) one gets $Q$, the initial state must be in the form
\begin{eqnarray}\label{defQminus1}
Q^-\!=\!\{n_1,\dots,n_g\!+\!1,\dots,n_l\!+\!1,\dots,n_{g+l}\!-\!1,\dots\}, 
\end{eqnarray}
when $g<l$, or  
\begin{eqnarray}\label{defQminus2}
Q^-=\{n_1,\dots,n_g\!+\!2,\dots,n_{2g}\!-\!1,\dots\},
\end{eqnarray}
for $g=l$. 

Now, the aim is to write the master equation for the probability, $W(Q,t)$, to find the coagulating system in the state $Q$ at the instant $t$. 
The equation has a known form:
\begin{eqnarray}\label{Master1}
\frac{dW(Q,t)}{dt}&=&\sum_{Q^-}A(Q^-\rightarrow Q)W(Q^-,t)\\\nonumber &-&\sum_{Q^+}A(Q\rightarrow Q^+)W(Q,t),
\end{eqnarray}
where $A(Q\rightarrow Q^+)$ is the probability per unit time to pass from the state $Q$ to $Q^+$. By definition, for a pair of particles of sizes $g$ and $l$, the rate of coagulation is $K(g,l)/V$, where $V$ represents volume of the system. Summed up over all such pairs, the transition rate from $Q$ to $Q^+$ is equal to:
\begin{eqnarray}\label{rateA}
A(Q\rightarrow Q^+)=\frac{K(g,l)}{2V}n_g(Q)\left(n_l(Q)-\delta_{g,l}\right),
\end{eqnarray}
where $\delta_{g,l}$ is the Kronecker delta. The expression for $A(Q^-\rightarrow Q)$ has a similar form as Eq.~(\ref{rateA}). 

The problem is, however, that the master equation in its original form, Eq.~(\ref{Master1}), is not easy to work with. Fortunately, the equation for the generating functional of $W(Q,t)$, which follows from  Eq.~(\ref{Master1}) is much more convenient. The mentioned functional is defined as 
\begin{equation}\label{defPsi0}
\Psi(X,t)=\sum_QW(Q,t)X^Q,
\end{equation}
where the notation:
\begin{equation}\label{defX}
X^Q=x_1^{n_1(Q)}x_2^{n_2(Q)}\dots x_g^{n_g(Q)}\dots,
\end{equation}
is employed. This functional, like the time-dependent probability distribution $W(Q,t)$, contains the complete information about the studied system. In particular, monodisperse initial condition corresponds to:
\begin{equation}\label{Psit0}
\Psi(X,0)=x_1^M,
\end{equation}
whereas the normalization of $W(Q,t)$ to unity, i.e. $\sum_QW(Q,t)=1$, corresponds to the condition:
\begin{equation}\label{NormPsi}
\Psi(X\!=\!1,t)=1,
\end{equation}
where $X\!=\!1$ means that $\forall_{g}\;x_g\!=\!1$. In addition, the following expression for the average number of $g$-mers, $\langle n_g\rangle=\sum_Qn_g(Q)W(Q,t)$, is true:
\begin{equation}\label{MeanNg0}
\langle n_g(t)\rangle=\left.\hat{n}_g\Psi(X,t)\right|_{X=1},
\end{equation}
where the occupation number operator is defined as:
\begin{equation}\label{hatNg0}
\hat{n}_g=x_g\frac{\partial}{\partial x_g}.
\end{equation}

So, to derive the equation for $\Psi(X,t)$ one just needs to multiply both sides of Eq.~(\ref{Master1}) by $X^Q$ and then sum it up over all states. Using, in the course of these transformations, the identities:
\begin{eqnarray}\nonumber
x_{g+l}\;\frac{\partial^2}{\partial x_g\partial x_l}X^Q&=&n_g(Q)\left(n_l(Q)-\delta_{g,l}\right) X^{Q^+},\\\nonumber x_gx_l\;\frac{\partial^2}{\partial x_g\partial x_l}X^Q&=&n_g(Q)\left(n_l(Q)-\delta_{g,l}\right) X^{Q},
\end{eqnarray}
in place of the master equation for $W(Q,t)$, one gets the following equation for $\Psi(X,t)$:
\begin{equation}\label{Master2}
V\frac{\partial\Psi}{\partial t}=\hat{\mathcal{L}}\Psi,
\end{equation}
where the evolution operator $\hat{\mathcal{L}}$ has the form:
\begin{equation}\label{hatL1}
\hat{\mathcal{L}}=\frac{1}{2}\sum_{g,l}K(g,l)(x_{g+l}-x_gx_l)
\frac{\partial^2}{\partial x_g\partial x_l}.
\end{equation}

The purpose of this contribution is to solve Eq.~(\ref{Master2}) for the product kernel under arbitrary initial conditions. To this aim, we use its solution for the same kernel and monodisperse initial conditions, that was first reported by Lushnikov in 2004 \cite{2004_PRLLushnikov}, and which, for the sake of completeness, is outlined later in this section.

\subsection{Product kernel and monodisperse initial conditions}

For the product kernel,
\begin{equation}
K(g,l)=2gl,
\end{equation}
the evolution operator, Eq.~(\ref{hatL1}), can be (after some algebra) rewritten as follows
\begin{equation}\label{hatL2}
\hat{\mathcal{L}}=\sum_{g,l}gl\;x_{g+l}\frac{\partial^2}{\partial x_g\partial x_l}+ \sum_gg^2\hat{n}_g-M^2.
\end{equation}
In Ref.~\cite{2004_PRLLushnikov}, it was shown that the solution to Eq.~(\ref{Master2}) with $\hat{\mathcal{L}}$ given by Eq.~(\ref{hatL2}) can be constructed in the form:
\begin{equation}\label{defPsi1}
\Psi(X,t)=M!\;\mathbf{Coef}\!\left(\!\frac{1}{z^{M+1}} \exp\left[\sum_{n=1}^\infty\!z^n a_n(t) x_n\right]\!\right)\!,
\end{equation}
where the $\mathbf{Coef}$ operation is defined as \cite{book_Egorychev}
\begin{equation}\label{defCoef}
\mathbf{Coef}\!\left(\sum_n b_nz^n\!\right)=b_{-1}.
\end{equation}
After substituting Eq.~(\ref{defPsi1}) into (\ref{MeanNg0}), the expression for the exact average number of clusters of size $g$ at time $t$ can be written as:
\begin{equation}\label{MeanNg1}
\langle n_g(t)\rangle=M!\;a_g(t)\;\mathbf{Coef}\left(\!\frac{z^g}{z^{M+1}} e^{G(z,t)}\!\right)\!,
\end{equation}
where $G(z,t)$ is the generating function for the sequence $a_g(t)$, that is:
\begin{equation}\label{defG0}
G(z,t)=\sum_{g=1}^\infty a_g(t)z^g.
\end{equation}

The Lushnikov achievement was that he calculated the coefficients $a_g(t)$ and their generating function $G(z,t)$ under the assumption of the product kernel and monodisperse initial conditions. (Note that, the mentioned conditions, Eq.~(\ref{Psit0}), can be recovered from Eq.~(\ref{defPsi1}) after putting $a_g(0)=\delta_{g,1}$.) Thereby, he indirectly determined the probability distribution $W(Q,t)$ underlying time evolution of the considered coagulating system, and he directly obtained its exact distribution of cluster sizes. In Sect.~\ref{SecAF}, we show how by using the Lushnikov result, one can determine these coefficients for arbitrary initial conditions. To this end, we need to explain in more detail the Lushnikov method, which we do below.

First, one can show that: If the functional~(\ref{defPsi1}) is the solution to Eq.~(\ref{Master2}), then its coefficients $a_g(t)$ satisfy the following equation:
\begin{equation}\label{Master3}
V\frac{da_g}{dt}=\sum_{l=1}^{g-1}l(g-l)a_la_{g-l}-Mga_g+g^2a_g.
\end{equation}
Using the generating function for these coefficients, Eq.~(\ref{defG0}), from Eq.~(\ref{Master3}) one gets the differential equation for $G(z,t)$:
\begin{equation}\label{Master3a}
V\frac{\partial G}{\partial t} =\left(z\frac{\partial G}{\partial z}\right)^2-Mz\frac{\partial G}{\partial z}+z\frac{\partial }{\partial z}z\frac{\partial G}{\partial z}.
\end{equation}
Then, as a result of the below substitution:
\begin{equation}\label{defD0}
G(z,t)=\log D(ze^{-Mt/V},t),
\end{equation}
Eq.~(\ref{Master3a}) can be further transformed into the linear equation for $D(z,t)$, that is:
\begin{equation}\label{Master4}
V\frac{\partial D}{\partial t}=z\frac{\partial }{\partial z}z\frac{\partial D}{\partial z}.
\end{equation}

Eq.~(\ref{Master4}) was solved by Lushnikow under the initial condition: $D(z,0)=e^z$, which corresponds to the monodisperse initial condition (cf.~Eqs.~(\ref{defG0}) and~(\ref{defD0}) for~$a_g(0)=\delta_{g,1}$) and provides:
\begin{equation}\label{defD1}
D(z,\tau)=\sum_{g=0}^\infty e^{g^2\tau}\frac{z^g}{g!},
\end{equation}
where $\tau=t/V$. Substituting the above result into Eq.~(\ref{defD0}) and then into~(\ref{MeanNg1}), Lushnikov found that, cf.~Eq.~(\ref{MeanNg1}),
\begin{equation}\label{defeG0}
\mathbf{Coef}\left(\!\frac{z^g}{z^{M+1}} e^{G(z,\tau)}\!\right)= \frac{1}{(M-g)!}e^{(g^2-Mg)\tau}.
\end{equation}
Next, using certain combinatorial identities, he expanded $\ln[D(z,\tau)]$, into a power series in $z$ and obtained, after some algebra, the strict formula for the coefficients $a_g(t)$:
\begin{equation}\label{ag1}
a_g(t)=\frac{1}{g!}e^{g\tau(1-M)}(e^{2\tau}-1)^{g-1}\textbf{F}_{g-1}(e^{2\tau}),
\end{equation}
where $\textbf{F}_g(x)$ are the so-called Mallows-Riordan polynomials \cite{1998_Knuth, 1998_Flojolet}. Finally, inserting Eqs.~(\ref{defeG0}) and~(\ref{ag1}) into Eq.~(\ref{MeanNg1}) Lushnikov got his main result - the exact average particle mass spectrum in coagulating systems evolving according to the product kernel under the monodisperse initial conditions:
\begin{equation}\label{MeanNg2}
\langle n_g(\tau)\rangle =\binom{M}{g} e^{(g^2-2Mg+g)\tau} (e^{2\tau}-1)^{g-1}\textbf{F}_{g-1}(e^{2\tau}).
\end{equation}

At this point, after a large dose of mathematics and before its next portion, to encourage the reader to read further, we would like to say that: The calculations presented further in this paper, although they refer to the results of this section, due to the introduction of the so-called Bell polynomials, are more concise and therefore easier. In addition, the Bell polynomials, which we use in our derivations, are much better known combinatorial creatures than the Mallows-Riordan polynomials, which appear in the Lushnikov solution. Recently, the Bell polynomials are employed in more and more papers in the field of theoretical physics (see e.g. \cite{2018_PREFronczak, book_Aldrovandi, 2012_PREFronczak, 2013_RepSiudem, 2014_RepFronczak, 2016_SciRepSiudem, 2018_JStatMechZhou}), which reveals their great usefulness and unexpected universality. Not without significance is also the fact that Bell polynomials, unlike the Mallow-Riordan polynomials, can be calculated using special built-in functions in different computing environments (including Wolfram Mathematica and Matlab).

Below we introduce Bell polynomials to the extent that is necessary to keep track the rest of our paper. 

\subsection{Bell polynomials}

There are two kinds of Bell polynomials \cite{book_Comtet, FaaDiBruno}: partial and complete, which are respectively given by:
\begin{eqnarray}\label{pBell0}
\!&\!&\!\textbf{B}_{M,m}\!=\!\textbf{B}_{M,m}(x_1,x_2,\!\dots)\;\;\;\;\;\;\;\mbox{and} \\\label{cBell0}
\!&\!&\!\textbf{Y}_{M}\!=\!\textbf{Y}_M(x_1,x_2,\!\dots)\!=\!\sum_{m=1}^M\!\textbf{B}_{M,m}(x_1,x_2,\!\dots).
\end{eqnarray}  
They are the polynomials in an infinite number of variables $\{x_n\}=x_1,x_2,\dots$ defined by the formal double series expansion:
\begin{eqnarray}\label{Bell0a}
\Phi(t,u)&=&\exp\left[u\sum_{n=1}^\infty \frac{t^n}{n!}x_n\right]\\\label{Bell0b}
&=&1+\sum_{M=1}^\infty\frac{t^M}{M!}\sum_{m=1}^M u^m\textbf{B}_{M,m}(\{x_n\})\\\label{Bell0c}
&=&1+\sum_{M=1}^\infty\frac{t^M}{M!}\;\textbf{Y}_M(\{ux_n\}).
\end{eqnarray}

The exact expression for partial Bell polynomials is the following:
\begin{eqnarray}\label{pBell1a}
\textbf{B}_{M,m}(\{x_n\})\!&=&\!M!\sum_{\{n_g\}}\frac{x_1^{n_1}x_2^{n_2}\dots}{n_1!n_2!\dots(1!)^{n_1}(2!)^{n_2}\dots}\;\;\;\\\label{pBell1b}\!&=&\!
M!\sum_{\{n_g\}}\prod_{g=1}^{M-m+1}\frac{1}{n_g!} \left(\frac{x_g}{g!}\right)^{\!\!n_g},
\end{eqnarray}
where the summation is taken over all non-negative integers $\{n_g\}$ that satisfy the below equations: 
\begin{equation}\label{BellX}
\sum_{g=1}^Mn_g=m,\;\;\;\mbox{and}\;\;\;\sum_{g=1}^Mg\,n_g=M.
\end{equation}

Although it is of minor importance in this paper, it may be interesting to know that these polynomials have an intuitive combinatorial meaning, which is easy to deduce from Eq.~(\ref{pBell1a}): Namely, when all $x_g$ are non-negative integers, $\textbf{B}_{M,m}(\{x_n\})$ gives the number of possible partitions of a set of $M$ elements (e.g. balls) into $m$ non-empty and disjoint subsets, assuming that each subset of size $g$ can be found in one of $x_g$ internal states. With this interpretation, in Eq.~(\ref{pBell1a}), the variables $\{n_g\}=n_1,n_2,\dots$ stand for the number of subsets of a given size.

Bell polynomials have a number of interesting properties, which are of use in the rest of this paper. For example, the derivative of $\textbf{B}_{M,m}(\{x_n\})$ with respect to $x_g$~is:
\begin{equation}\label{pBell2}
\frac{d\textbf{B}_{M,m}(\{x_n\})}{dx_g}=\binom{M}{g}\textbf{B}_{M-g,m-1} (\{x_n\}).
\end{equation}
Accordingly, from Eq.~(\ref{cBell0}) one has:
\begin{equation}\label{cBell2}
\frac{d\textbf{Y}_M(\{x_n\})}{dx_g}=\binom{M}{g}\textbf{Y}_{M-g}(\{x_n\}).
\end{equation}
Another important identity states the following:
\begin{equation}\label{pBell3}
\textbf{B}_{M,m}(\{a^n b\;x_n\})=a^Mb^m\;\textbf{B}_{M,m}(\{x_n\}).
\end{equation}
Of great importance is also the below expression with the use of Bell polynomials:
\begin{equation}\label{logBell0}
\log\left[1+\sum_{n=1}^\infty \frac{t^n}{n!}x_n \right] =\sum_{g=1}^\infty \textbf{L}_g(\{x_n\})\frac{t^g}{g!},
\end{equation}
where $\textbf{L}_g$ are the so-called logarithmic polynomials which are defined as
\begin{equation}\label{logBell1}
\textbf{L}_g(\{x_n\})=\sum_{k=1}^g(-1)^{k-1}(k-1)!\;\textbf{B}_{g,k}(\{x_n\}).
\end{equation}
Other identities for Bell polynomials will be revealed successively as they are used.

\section{Product kernel and arbitrary initial conditions} \label{SecAF}

\subsection{Basic equations}

To start with, we note that the functional (\ref{defPsi1}), which was proposed by Lushnikov as the general solution to Eqs.~(\ref{Master2})-(\ref{hatL1}), is equivalent to the complete Bell polynomial:
\begin{equation}\label{PsiBell0}
\Psi(X,t)=\textbf{Y}_M\left(\{n!a_n(t)x_n\}\right).
\end{equation}
This equivalence follows from Eqs.~(\ref{Bell0a})-(\ref{Bell0c}) and has some consequences, the most important of which is that it enforces a specific form of the initial conditions:
\begin{equation}\label{PsiBellt0}
\Psi(X,0)=\textbf{Y}_M\left(\{n!a_n(0)x_n\}\right).
\end{equation}
(The truth is that not all the initial conditions can be written in the form of Bell polynomials, even if they are consistent with the definition~(\ref{defPsi0}) and normalized~(\ref{Psit0}). We will return to these issues later in this section.)

For arbitrary initial conditions, the generating function for the sequence $\{a_n(0)\}$ has the following form, Eq.~(\ref{defGt0}):
\begin{equation}\label{defGt0}
G(z,0)=\sum_{g=1}^\infty a_g(0)z^g.
\end{equation}
Accordingly, at $t\!=\!0$, the auxiliary function $D(z,0)$, Eq.~(\ref{defD0}), can be written as:
\begin{eqnarray}\label{defDt0}
D(z,0)=e^{G(z,0)}\stackrel{\small{Eq.(\ref{Bell0c})}}{=}1+\sum_{g=1}^\infty b_g\frac{z^g}{g!},
\end{eqnarray}
where 
\begin{equation}\label{bn}
b_g=\textbf{Y}_g(\{n!a_n(0)\}).
\end{equation}
Now, since the differential equation for $D(z,t)$, Eq.~(\ref{Master4}), is linear, one can use its solution for the monodisperse initial conditions, Eq.~(\ref{defD1}), to write the corresponding solution for arbitrary initial conditions:
\begin{eqnarray}\label{defDt}
D(z,\tau)=1+\sum_{g=1}^\infty b_g\;e^{g^2\tau}\;\frac{z^g}{g!},
\end{eqnarray}
where $\tau=t/V$. From the above result one also has, cf.~Eq.~(\ref{defD0}),
\begin{eqnarray}\label{defGt}
G(z,\tau)=\log\left[1+\sum_{g=1}^\infty c_g\frac{z^g}{g!}\right],
\end{eqnarray}
where
\begin{equation}\label{cn}
c_g=b_ge^{(g^2-gM)\tau}.
\end{equation}

The obtained functions, $D(z,\tau)$ and $G(z,\tau)$, can now be used to complete derivation of the exact expression for the average number of clusters of size $g$ at time $t$. First, from Eq.~(\ref{MeanNg1}), by using~(\ref{defGt}) one gets:
\begin{eqnarray}\label{MeanNg4a}
\langle n_g(t)\rangle\!&=&\!\frac{M!}{(M\!-\!g)!}\; a_g(t)\;c_{M\!-\!g} \\\nonumber\!&=&\!\frac{M!}{(M\!-\!g)!} \;\textbf{Y}_{M\!-\!g}(\{n!a_n(0)\})\;a_g(t)\;e^{(g^2\!-\!gM)\tau}\!.
\end{eqnarray}
Second, when expanding Eq.~(\ref{defGt}) in power series of $z$, with the help of logarithmic polynomials, Eq.~(\ref{logBell0})-(\ref{logBell1}), one obtains: 
\begin{equation}
G(z,t)=\sum_{g=1}^\infty\textbf{L}_g(\{c_n\})\frac{z^g}{g!},
\end{equation}
and finally:
\begin{equation}\label{agt}
a_g(t)=\frac{1}{g!}\textbf{L}_g(\{c_n\}).
\end{equation}

Eq.~(\ref{MeanNg4a}), supplemented with Eq.~(\ref{agt}), is the most important result of this paper. 

\subsection{Examples}

\subsubsection{Monodisperse initial conditions}

The well-known property of partial Bell polynomials states that: 
\begin{equation}\label{pomBell0}
\textbf{B}_{M,m}(\{\delta_{g,n}x_n\})=\delta_{M,gm}\;\frac{M!}{m!(g!)^m}\;x_g^m.
\end{equation}
For $g\!=\!1$ the above identity simplifies to:
\begin{equation}\label{pomBell1}
\textbf{B}_{M,m}(x_1,0,0,\dots)=\delta_{M,m}\;x_1^m,
\end{equation}
which, according to Eq.~(\ref{cBell0}), gives
\begin{equation}\label{pomBell2}
\textbf{Y}_M(x_1,0,0,\dots)=x_1^M.
\end{equation}

In the case of monodisperse initial conditions, the sequence $\{a_n(0)\}$ is defined by:
\begin{equation}\label{agt01}
\forall_{g}a_g(0)=\delta_{g,1}, 
\end{equation}
which leads to:
\begin{equation}\label{Yt0}
\forall_{g} \textbf{Y}_g(\{n!a_n(0)\})=1. 
\end{equation}

With the use of Eq.~(\ref{Yt0}), the coefficients: $b_g$, Eq.~(\ref{bn}), and $c_g$, Eq.~(\ref{cn}), are respectively given by:
\begin{equation}\label{bn1}
b_g=1,\;\;\;\;\;\mbox{and}\;\;\;\;\;c_g=e^{(g^2-gM)\tau}.
\end{equation}
Inserting them, first into Eq.~(\ref{agt}), and then into Eq.~(\ref{MeanNg4a}), one gets:
\begin{equation}\label{agt1}
a_g(t)\stackrel{\small{Eq.(\ref{pBell3})}}{=}\; \frac{e^{-gM\tau}}{g!}\textbf{L}_g(\{e^{n^2\tau}\}),
\end{equation}
and finally:
\begin{equation}\label{Ngt1}
\langle n_g(t)\rangle=\binom{M}{g}e^{(g^2-2gM)\tau} \textbf{L}_g(\{e^{n^2\tau}\}).
\end{equation}

Eq.~(\ref{Ngt1}) is equivalent to Eq.~(\ref{MeanNg2}), which was first derived by Lushnikov in 2004. Strict proof of this equivalence goes beyond the scope of this paper, the more that, there is a lack of contributions in which interrelationships between Mallows-Riordan and Bell polynomials would be discussed. Nevertheless, direct correspondence between Eqs.~(\ref{Ngt1}) and~(\ref{MeanNg2}) has been confirmed in numerical simulations, see Fig.~1. 

\begin{figure}[h]
\includegraphics[scale=0.57]{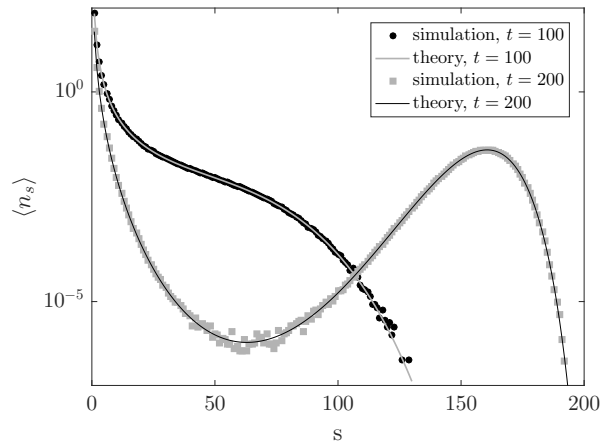}
\caption{The average number of clusters $\langle n_s \rangle$ for monodisperse initial conditions. The circles and squares correspond to the results obtained by simulation. The lines correspond to the solution given by Eq. (\ref{Ngt1}). The parameters of the studied system are $M=200$, $t=100$ for black circles and $M=200$, $t=200$ for gray squares. Lushnikov's theoretical prediction strictly overlaps with our prediction. Prediction fits simulation perfectly. The simulation results are averaged over $2.5\times10^6$ independent runs for $t=100$ and over $10^7$ independent runs for $t=200$ to obtain smoother statistics for lower values of $\langle n_s \rangle$.}
\end{figure}

\subsubsection{Other homogeneous initial conditions\\(dimers, trimers, $\dots$)}

When the system begins to evolve from $k-$mers only (eg. dimers, trimers, $\dots$), the natural choice for the sequence $\{a_n(0)\}$ is:
\begin{equation}\label{agt0n}
\forall_{g}a_g(0)=\delta_{g,k}. 
\end{equation}
For this choice, however, except for $k\!=\!1$ (monodisperse conditions), the initial functional $\Psi(X,0)$, Eq.~(\ref{PsiBellt0}), does not satisfy the normalization condition, cf.~Eq.~(\ref{NormPsi}):
\begin{equation}\label{Psit0n}
\Psi(X\!=\!1,0)=\frac{M!}{(M/k)!}\neq 1,
\end{equation}
where the following identity was used:
\begin{equation}\label{pomBell3}
\textbf{Y}_{g}\left(\{\delta_{n,k}n!\}\right)\! \stackrel{\small{Eq.(\ref{pomBell0})}}{=}\frac{g!}{(g/k)!}\; [g\mbox{\;mod\;}k=0],
\end{equation}
with $[P]$ standing for the Iverson bracket, which converts the logical proposition $P$ into a number that is equal to~$1$ if the proposition is satisfied, and~$0$ otherwise. 

Because of Eq.~(\ref{Psit0n}), resulting from the master equation, the functional $\Psi(X,t)$, Eq.~(\ref{PsiBell0}), also does not meet normalization. 
To cope with this problem, it is enough to divide $\Psi(X,t)$ by $\Psi(1,0)=const$. To justify this treatment, one just notes that $\Psi(X,t)$ given by Eq.~(\ref{defPsi1}) was ``proposed'' in a rather arbitrary way. Looking at Eq.~(\ref{Master2}), from which Eq.~(\ref{Master3}) is derived, one can see that multiplying (dividing) $\Psi(X,t)$ by a constant does not affect the coefficients $a_g(t)$. That is, to describe coagulation with product kernel under homogeneous initial conditions, Eq.~(\ref{agt0n}), instead of Eq.~(\ref{PsiBell0}), one ``can'' use:
\begin{eqnarray}\label{Psitn}
\Psi(X,t)=\frac{\textbf{Y}_M\left(\{n!a_n(t)x_n\}\right)} {\textbf{Y}_M\left(\{n!a_n(0)\}\right)}.
\end{eqnarray}
In accordance with Eq.~(\ref{Psitn}), however, from Eq.~(\ref{MeanNg0}) it also follows that, instead of (\ref{MeanNg4a}), one ``should'' use the below expression: 
\begin{eqnarray}\label{Ngna}
\langle n_g(t)\rangle\!&=&\!\frac{M!}{(M\!-\!g)!}\frac{a_g(t)\;c_{M\!-\!g}} {\textbf{Y}_{M}(\{n!a_n(0)\})},
\end{eqnarray}
with $a_g(t)$ still given by Eq.~(\ref{agt}).

Using Eq.~(\ref{pomBell3}) in Eq.~(\ref{Ngna}), one gets the expression for the exact average number of clusters of size $g$ at time $t$, when the coagulating system starts to evolve from $k$-mers only:
\begin{align}\nonumber
\langle n_g(t)\rangle=& \frac{\left(\frac{M}{k}\right)!}{\left(\frac{M-g}{k}\right)!g!}e^{(g^2-2gM)\tau}\times\\\label{Ngt2} &\textbf{L}_g\left(\left\{\frac{n!}{\left(\frac{n}{k}\right)!} e^{n^2\tau} [P1]\right\}\right)[P2],
\end{align}
where the conditions $P1$ and $P2$ are respectively given by:
$n\;\mbox{mod}\;k=0$ and $(M\!-\!g)\;\mbox{mod}\;k=0$. For $k\!=\!1$, the above result turns into the just derived formula for the monodisperse initial conditions, Eq.~(\ref{Ngt1}).
\newline
\begin{figure}[h]
\includegraphics[scale=0.57]{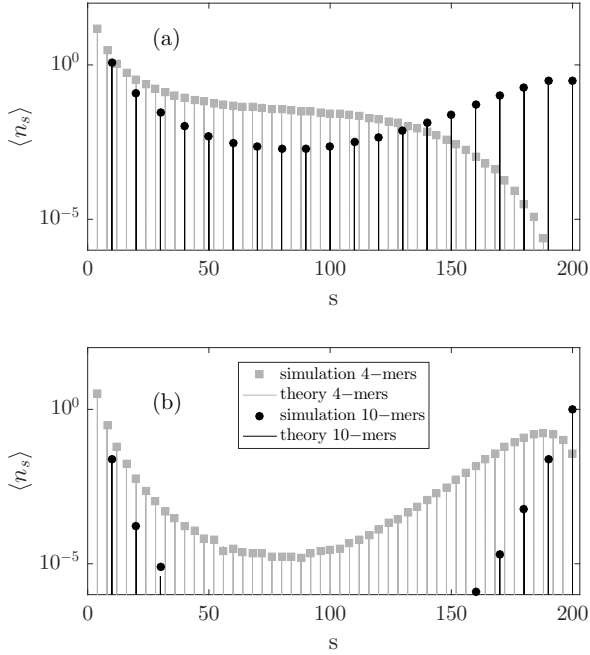}
\caption{Numerical simulations vs Eq.~(\ref{Ngt2}). The system studied consisted initially of $M=200$ monomeric units grouped into $4-$mers only (gray squares) or $10-$mers only (black circles). In a) one can see the behaviour of the systems at the beginning of the coalescence for $t=30$. In b) one can see the later stage of the process for $t=70$. Both in simulations and theoretical predictions (solid lines) not all of the cluster sizes are occupied but only those which are multiples of the initial cluster sizes. Theoretical prediction reproduces the simulation result perfectly. The simulation results are averaged over $2.5\times10^6$ independent runs}
\end{figure}

\subsubsection{A mixture of monomers and dimers}\label{SecAF23}

In what follows we study kinetics of the coagulating system with product kernel under the initial conditions in which the system consists of monomers and dimers in a given proportion ratio. We start with an important observation, which is not obvious at first glance, that the general definition of the functional $\Psi(X,t)$, Eq.~(\ref{Psitn}), enforces a specific form of the initial functional, $\Psi(X,0)$. In particular, $\Psi(X,0)=\frac{1}{2}x_1^M+ \frac{1}{2}x_2^{M/2}$ (which states: with equal probability the system starts to evolve from only monomers or only dimers) does not belong to the class that is defined by Eq.~(\ref{Psitn}). In fact, using the approach described in this paper, one can only examine a kind of mixed initial conditions in which, at time $t=0$, the system consists of monomers and dimers of various predefined mean concentrations. 

To explain the nature of this ``mixture'', let us assume that the initial sequence $\{a_n(0)\}$ is defined as:
\begin{equation}\label{an0_12a}
\{a_n(0)\}=a_1(0),a_2(0),0,0,0,\dots,
\end{equation}
where $a_1(0)$ and $a_2(0)$ are some non-negative parameters that determine the average initial values of $\langle n_1(0)\rangle$ and $\langle n_2(0)\rangle$.
At this point, let us note that for the considered initial conditions, according to Eq.~(\ref{Ngna}), one has: $\langle n_g(0)\rangle=0$ for $g>2$. (In what follows, to avoid too-extensive notation referring to the initial values $a_g(0)$ and $\langle n_g(0)\rangle$, we omit to explicitly write the time-dependence and simply write $a_g^0$ and $\langle n_g^0\rangle$.)

To show, how $\langle n_1^0\rangle$ and $\langle n_2^0\rangle$ depend on $a_1^0$ and $a_2^0$ note that from Eq.~(\ref{Ngna}) the following expressions arise:
\begin{equation}\label{md1}
\frac{\langle n_1^0\rangle_{_{\!M}}}{\langle n_2^0\rangle_{_{\!M}}}=\frac{a_1^0}{a_2^0}\textbf{H}(\{a_n^0\}),
\end{equation}
and
\begin{equation}\label{md2}
\frac{\langle n_1^0\rangle_{_{\!M\!-\!1}}}{a_1^0}=\frac{1}{\textbf{H}(\{a_n^0\})},
\end{equation}
where $\langle n_g^0\rangle_{_M}$ and $\langle n_g^0\rangle_{_{M\!-\!1}}$ stand for the average initial number of clusters of size $g$ in systems of size $M$ and $M\!-\!1$, respectively, both obtained under the same initial conditions, and
\begin{equation}\label{md3}
\textbf{H}(\{a_n^0\})=\frac{\textbf{Y}_{M-1}(\{a_n^0\})} {\textbf{Y}_{M-2}(\{a_n^0\})} \frac{(M\!-\!2)!}{(M\!-\!1)!}.
\end{equation}
Substituting Eq.~(\ref{md2}) into~(\ref{md1}), and then assuming that
\begin{equation}\label{cond0}
\langle n_1^0\rangle_{_{\!M}}\!=\!\langle n_1^0\rangle_{_{\!M\!-\!1}}\!\equiv\! \langle n_1^0\rangle\;\;\;\;\;\mbox{and}\;\;\;\;\; \langle n_2^0\rangle_{_{\!M}}\!\equiv\!\langle n_2^0\rangle,
\end{equation}
one gets:
\begin{equation}\label{cond1}
\frac{\langle n_1^0\rangle^2}{\langle n_2^0\rangle}=\frac{(a_1^0)^{\;2}}{a_2^0}\equiv a.
\end{equation}

Now, taking into account that, cf~Eq.~(\ref{ngEq}): 
\begin{equation}
\langle n_1^0\rangle+2\langle n_2^0\rangle=M,
\end{equation}
one can show that for a given value of the parameter~`$a$', Eq.~(\ref{cond1}), the average initial number of monomers is:
\begin{equation}
\langle n_1^0\rangle=\frac{\sqrt{a^2+8Ma}-a}{4}.
\end{equation}
Furthermore, the initial sequence (\ref{an0_12a}) can be rewritten in the convenient form:
\begin{eqnarray}\label{an0_12b}
\{a_n^0\}=\langle n_1^0\rangle,\langle n_2^0\rangle,0,0,0,\dots.
\end{eqnarray}

Finally, inserting (\ref{an0_12b}) into Eq.~(\ref{Ngna}), and using the well-known identity for Bell polynomials (which is given further in the text), one gets:
\begin{equation}\label{Ng_an12a}
\langle n_g(t)\rangle=\binom{M}{g}e^{(g^2-2gM)\tau} \textbf{L}_g(\{e^{n^2\tau}K_n\})\frac{K_{_{M\!-\!g}}}{K_{_M}},
\end{equation}
where
\begin{equation}\label{Ng_an12b}
K_n\equiv K_n(a)=\sum_{k=\lceil\frac{n}{2}\rceil}^{n}\frac{n!} {(n\!-\!k)!(2k\!-\!n)!}a^k.
\end{equation}

To obtain the result (\ref{Ng_an12a}) the following identity for Bell polynomials was used:
\begin{align}
\textbf{B}_{M,m}&(x_1,2x_2,0,0,\dots)\\\nonumber=& \sum_{l=0}^m\binom{M}{l}x_1^l\;\textbf{B}_{M-l,m-l}(0,2x_2,0,0,\dots)\\ \nonumber=& \sum_{l=0}^m\frac{M!}{(M\!-\!m)!l!}x_1^l\;\textbf{B}_{M-m,m-l}(x_2,0,0,\dots) \\\nonumber\stackrel{\small{Eq.(\ref{pomBell1})}}{=}& 
\sum_{l=0}^m\frac{M!}{(M\!-\!m)!l!}\;x_1^lx_2^{m-l}\delta_{M-m,m-l}\\
\nonumber=& 
\frac{M!}{(M\!-\!k)!(2k!-\!M)!}\left(\!\frac{x_1^2}{x_2}\!\right)^{\!\!m}\!\left(\!\frac{x_2}{x_1}\!\right)^{\!\!M}
\left[m\in\left\langle\lceil M/2\rceil,M\right\rangle\right]
\end{align}

\begin{figure}[h]
\includegraphics[scale=0.57]{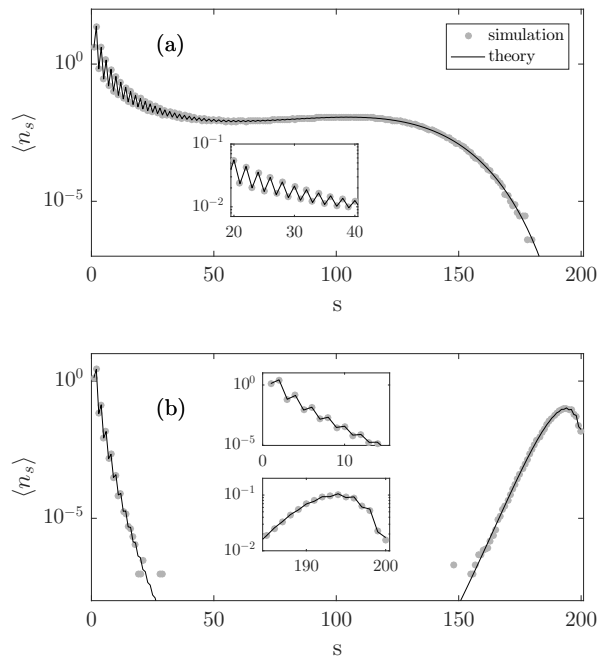}
\caption{Numerical simulations compared with solutions given by Eq.~(\ref{Ng_an12a}) for mixed initial conditions. The system initially consisted of $M=200$ monomeric units grouped into 8 monomers and 96 dimers. a) For $t=70$ one can observe that the plot is not smooth as the occurrence of odd-size clusters is less probable than even-size ones due to the initial conditions. b) For the later stage of the coalescence, $t=180$, the lack of equiprobability for even- and odd-cluster sizes is preserved. As can be seen in insets at all stages of the process theory fits simulation ideally and reproduces the saw-like serration of $\langle n_s \rangle$. The simulation results were averaged over $2.5\times10^6$ independent runs for (a) and $10^7$ for (b).}
\end{figure}


For the simulation algorithm that we use to produce numerical simulation data, see Supplemental Material \cite{SupMat}. We also describe there the issues of calculating theoretical predictions and the arbitrary precision library used for these calculations.


\section{Concluding remarks}\label{SecSum}

With regard to exact results, there is always a question: To what extent they can be extended and used for studying other systems? And although answers to such questions are often wishful thinking, below we give some suggestions on what can be done further that would have a measurable effect on the theory of coagulation. 

First, it would be interesting to apply the approach to study the coagulation process with product kernel and other initial conditions. For sure, of particular interest are exponentially and algebraically decaying initial mass spectra (i.e.~$\langle n_g^0\rangle\propto e^{-\lambda g}$ and~$\langle n_g^0\rangle\propto g^{-\alpha}$, respectively; the abbreviated notation of Sec.~\ref{SecAF23} is used here). Confirmation or falsification of the mean-field results obtained from the Smoluchowski equation \cite{2015_PhysALeyvraz}, related to non-trivial behavior of these systems in the vicinity of critical points, would be an important result. For those who would like to follow this suggestion, a valuable guideline may be that, for arbitrary initial cluster size distribution, the initial sequence $\{a_n^0\}$ can be written in the form analogous to Eq.~(\ref{an0_12b}):
$\{a_n^0\}=\{\langle n_g^0\rangle\}$. Theoretically, after substituting this sequence into Eq.~(\ref{Ngna}) one should get the result. Practically, however, it appears that the real challenge may arise as to: how to simplify the results obtained and how to find their asymptotic behavior. Fortunately, Bell polynomials have many useful properties \cite{book_Comtet, FaaDiBruno} that can help to solve this difficult task.

The second suggestion is that other kernels (not just the multiplicative one) under arbitrary initial conditions can likely be solved in the same way as shown in this paper. In view of the announced sensitivity of the coagulation process to the initial conditions \cite{2004_Menon}, such results would be of great importance for the theory of non-equilibrium phase transitions. The above is all the more feasible to do that the both kernels (additive and constant) under monodisperse initial conditions has been already solved using the Marcus-Lushnikov approach (see~\cite{1978_Lushnikov,2011_JPhysALushnikov}). 

\acknowledgments
This work has been supported by the National Science Centre of Poland (Narodowe Centrum Nauki, NCN) under grant no.~2015/18/E/ST2/00560 (A.F. and M.\L.).


\begin{thebibliography}{99}
	
	
	\bibitem{2006book_Seinfeld} J.H. Seinfeld, S. Pandis, \textit{Physics and Chemistry of the Atmosphere} (Wiley, Hoboken, NJ, 2006).
	
	\bibitem{2000book_Friedlander} S.K. Friedlander, \textit{Smoke, Dust and Haze} (Oxford University Press, Oxford, 2000).
	 
	\bibitem{1998book_Meakin} P. Meakin, \textit{Fractals, Scaling and Growth Far from Equilibrium} Cambridge Nonlinear Science Vol.~5 (Cambridge University Press, Cambridge, UK, 1998).
	
	
	\bibitem{1916_Smoluchowski} M.~Smoluchowski, \textit{Drei vortr\"{a}ge \"{u}ber diffusion bewegung und koagulation von kolloidteilchen}, Phys.~Z. \textbf{17}, 557-585 (1916). 

	\bibitem{1941a_Flory} P.~Flory, \textit{Molecular size distribution in three dimensional polymers. I. Gelation}, J.~Am.~Chem.~Soc.~\textbf{63}(11), 3083 (1941). 

	\bibitem{1941b_Flory} P.~Flory, \textit{Molecular size distribution in three dimensional polymers. II. Trifunctional branching units}, J.~Am.~Chem.~Soc.~\textbf{63}(11), 3091 (1941). 
	
	\bibitem{1941c_Flory} P.~Flory, \textit{Molecular size distribution in three dimensional polymers. III. Tetrafunctional branching units}, J.~Am.~Chem.~Soc.~\textbf{63}(11), 3096 (1941). 

	\bibitem{1943_Stockmayer} W.H.~Stockmayer, \textit{Theory of molecular size distribution and gel formation in branched gell polymers}, J.~Chem.~Phys. \textbf{11}, 45 (1943). 


	\bibitem{1968_Marcus} A.H. Marcus, \textit{Stochastic coallescence}, Technometrics \textbf{10}, 133-143 (1968).
	
	\bibitem{1978_Lushnikov} A.A. Lushnikov, \textit{Coaulation in finite systems}, J. Colloid Interface Sci. \textbf{65}, 276-285 (1978).
	
	\bibitem{1980_Ziff} R.M. Ziff, G. Stell, \textit{Kinetics of polymer gelation}, J. Chem. Phys. \textbf{73}, 3492-3499 (1980).
	
	\bibitem{1983_Ziff} R.M. Ziff, M.H. Ernst, E.M. Hendriks, \textit{Kinetics of gelation and universality}, J. Phys. A: Math. Gen. \textbf{16}, 2293 (1983).
	
	\bibitem{1983_Hendriks} E.M. Hendriks, M.H. Ernst, R.M. Ziff, \textit{Coagulation equation with gelation}, J. Stat. Phys. \textbf{31}, 519 (1983). 
	
	\bibitem{1986_Dongen} P.G.J. van Dongen, M.H. Ernst, \textit{On the occurrence of a gelation transition in Smoluchowski's coagulation equation}, J. Stat. Phys. \textbf{44}, 785 (1986).
	
	\bibitem{1999_RevAldous} D.J. Aldous, \textit{Deterministic and stochastic models for coalescence (aggregation and coagulation): a review of the mean field theory for probabilists}, Bernoulli \textbf{5}, 3 (1999).
	
	\bibitem{2003_PhysRepLeyvraz} F. Leyvraz, \textit{Scaling theory and exactly solved models in kinetics of irreversible aggregation}, Phys. Rep. \textbf{383}, 95 (2003).
	
	\bibitem{2006_PhysDWattis} J.A.D. Wattis, \textit{An introduction to mathematical models of coagulation-fragmentation processes: A discrete deterministic mean-field approach}, Physica D \textbf{222}, 1 (2006).
	
	\bibitem{2005_JPhysALushnikov} A.A.Lushnikov, \textit{Time evolution of a random graph}, J. Phys. A \textbf{38}, L777 (2005).
	
	\bibitem{2009_ScienceAchlioptas} D. Achlioptas, R.M. DSouza, J. Spencer, \textit{Explosive percolation in random networks}, Science \textbf{323}, 1453 (2009).
	
	\bibitem{2010_PRLCosta} R.A. da Costa, S.N. Dorogovtsev, A.V. Goltsev, J.F.F. Mendes, \textit{Explosive percolation transition is actually continuous}, Phys. Rev. Lett. \textbf{105}, 255701 (2010).
	
	\bibitem{2010_PRECho} Y.S. Cho, B. Kahng, D. Kim, \textit{Cluster aggregation model for discontinuous percolation transition}, Phys. Rev. E \textbf{81}, 030103(R) (2010).
	
	\bibitem{2016_PRLCho} Y.S. Cho, J.S. Lee, H.J. Hermann, B. Kahng, \textit{Hybrid percolation transition in cluster merging processes: Continuous varying exponents}, Phys. Rev. Lett. \textbf{116}, 025701 (2016).
	
	
	\bibitem{2016_Conv} O. Riordan, L. Wanke, \textit{Convergence of Achlioptas
processes via differential equations with unique solutions}, Combinatorics, Probability and Computing \textbf{25}, 154-171 (2016).
	
	\bibitem{2005book_Hein} J. Hein, M.H. Schierup, C. Wiuf, \textit{Gene Genealogies, Variation and Evolution. A Primer in Coalescent Theory} (Oxford University Press, New York, 2005).
	
	\bibitem{2014_PREMatsoukas} T. Matsoukas, \textit{Statistical thermodynamics of clustered populations}, Phys. Rev. E \textbf{90}, 022113 (2014).
	
	\bibitem{2014_SciRepMatsoukas} T. Matsoukas, \textit{Statistical thermodynamics of irreversible aggregation: the sol-gel transition}, Sci. Rep. \textbf{5}, 8855 (2014).
		
	\bibitem{2007_EcolModelSaadi} N. El Saadi, A. Bah, \textit{An individual-based model for studying the aggregation behavior in phytoplankton}, Ecol. Model. \textbf{204}, 193 (2007).
	
	\bibitem{2002_PREDubovik} V.M. Dubovik, A.G. Galperin, V.S. Richvitsky, A.A. Lushnikov, \textit{Analytical kinetics of clustering processes with cooperative action of aggregation and fragmentation}, Phys. Rev. E \textbf{66}, 016110 (2002).	
		
	\bibitem{2010book_Krapivsky} P.L. Krapivsky, S. Redner, E. Ben-Naim, \textit{A Kinetic View of Statistical Physics} (Chap. 5), New York, Cambridge University Press, 2010.	

	\bibitem{2012_PRELushnikov} A.A. Lushnikov, \textit{Supersingular mass distributions in gelling systems}, Phys. Rev. E \textbf{86}, 051139 (2012).	

	\bibitem{2013_PRELushnikov} A.A. Lushnikov, \textit{Postcritical behavior of a gelling system}, Phys. Rev. E \textbf{88}, 052120 (2013).	

	\bibitem{2018_PREFronczak} A. Fronczak, A. Chmiel, P. Fronczak, \textit{Exact combinatorial approach to finite coagulating systems}, Phys. Rev. E \textbf{97}, 022126 (2018).	

\bibitem{1974_Bayewitz} M.H. Bayewitz, J. Yerushalmi, S. Katz, R. Shinnar, \textit{The extent of correlations in a stochastic coalescence process}, J. Atmos. Sci. \textbf{31}, 1604-1614 (1974).

\bibitem{1985_Hendriks} E.M. Hendriks, J.L. Spouge, M. Eibl, M. Schreckenberg, \textit{Exact solutions for random coagulation processes}, Z. Phys. B \textbf{58}, 219 (1985).

\bibitem{2004_PRLLushnikov} A.A. Lushnikov, \textit{From sol to gel exactly}, Phys. Rev. Lett. \textbf{93}, 198302 (2004).

\bibitem{2005_PRELushnikov} A.A. Lushnikov, \textit{Exact kinetics of the sol-gel transition}, Phys. Rev. E \textbf{71}, 046129 (2005).

\bibitem{2011_JPhysALushnikov} A.A. Lushnikov, \textit{Exact kinetics of a coagulating system with the kernel $K=1$}, J. Phys. A \textbf{44}, 335001 (6pp) (2011).

\bibitem{2006_PhysDLushnikov} A.A. Lushnikov, \textit{Gelation in coagulation systems}, Physica D \textbf{222}, 37-53 (2006).

\bibitem{2011_NucLushnikov} A.A. Lushnikov, \textit{Field-theory methods in coagulation theory}, Phys. Atom. Nucl. \textbf{74}, 1096 (2011).	

\bibitem{2006_PhysDLeyvraz} F. Leyvraz, \textit{Scaling theory for gelling systems: Work in progress}, Physica D \textbf{222}, 21 (2006).

\bibitem{2015_PhysALeyvraz} F. Leyvraz, A.A. Lushnikov, \textit{Scaling anomalies in the sol-gel transition}, J. Phys. A \textbf{48}, 205002 (22p) (2015).

\bibitem{2004_Menon} G. Menon, R.L. Pego, \textit{Approach to self-similarity in Smoluchowski's coagulation equations}, Commun. Pure Appl. Math. \textbf{57}, 1197-1232 (2004).

\bibitem{book_Egorychev} G.P. Egorychev, \textit{Integral Representation and the Computation of Combinatorial Sums}, Transl. of Math. Monography, vol. \textbf{59}, Amer. Math. Soc., Providence, 1989. 

\bibitem{1998_Knuth} D.E. Knuth, \textit{Linear probing and graphs}, Algorithmica \textbf{22}, 561 (1998).

\bibitem{1998_Flojolet} P. Flojolet, P. Poblete, A. Viola, \textit{On the analysis of linear probing hushing}, Algorithmica \textbf{22}, 490 (1998).

\bibitem{book_Aldrovandi} R. Aldrovandi, \textit{Special Matrices of Mathematical Physics. Stochastic, Circulant and Bell Matrices}, World Scientific, Singapore, 2001.

\bibitem{2012_PREFronczak} A. Fronczak, \textit{The microscopic meaning of grand potential resulting from combinatorial approach to a general system of particles}, Phys. Rev. E \textbf{86}, 041139 (2012).

\bibitem{2013_RepSiudem} G. Siudem, \textit{Partition function of the model of perfect gas of clusters for interacting fluids}, Rep. Math. Phys. \textbf{72}, 85 (2013).

\bibitem{2014_RepFronczak} A. Fronczak, P. Fronczak, \textit{Exact expression for the number of energy states in lattice models}, Rep. Math. Phys. \textbf{73}, 1 (2014).

\bibitem{2016_SciRepSiudem} G. Siudem, A. Fronczak, P. Fronczak, \textit{Exact low-temperature series expansion for the partition function of the zero-field Ising model on the infinite square lattice}, Sci. Rep. \textbf{6}, 33523 (2016).

	
\bibitem{2018_JStatMechZhou}Chi-Chun Zhou, Wu-Sheng Dai, \textit{Canonical partition functions: ideal quantum gases, interacting classical gases, and interacting quantum gases}, J. Stat. Mech. 023105 (2018).
	
\bibitem{book_Comtet} L. Comtet, \textit{Advanced Combinatorics: The Art of Finite and Infinite Expansions} (Chap. 3.3), Reidel Publishing Company, Dordrecht, 1974.
	
\bibitem{FaaDiBruno} W.P. Johnson, \textit{The curious history of the Fa\`{a} di Brunos formula}, Am. Math. Mon. \textbf{109}, 217234 (2002).

\bibitem{SupMat} See Supplemental Material at [URL will be inserted by publisher] for numerical simulation algorithm and issues of theoretical calculations.

 
		
\end{thebibliography}
\end{document}